# Strongly-coupled quantum critical point in an all-in-all-out antiferromagnet

Yishu Wang[1], T. F. Rosenbaum[1], A. Palmer[2], Y. Ren[3], J.-W. Kim[3], D. Mandrus[4,5], Yejun Feng[1,6]

1) Division of Physics, Mathematics, and Astronomy, California Institute of Technology, Pasadena, California 91125, USA
2) The James Franck Institute and Department of Physics, The University of Chicago, Chicago, Illinois 60637, USA
3) The Advanced Photon Source, Argonne National Laboratory, Argonne, Illinois 60439, USA
4) Department of Materials Science and Engineering, University of Tennessee, Knoxville, Tennessee 37996, USA
5) Materials Science and Technology Division, Oak Ridge National Laboratory, Oak Ridge, TN 37831, USA
6) Okinawa Institute of Science and Technology Graduate University, Onna, Okinawa 904-0495, Japan

**Abstract:**
Dimensionality and symmetry play deterministic roles in the laws of Nature. They are important tools to characterize and understand quantum phase transitions, especially in the limit of strong correlations between spin, orbit, charge, and structural degrees of freedom. Using newly-developed, high-pressure resonant x-ray magnetic and charge diffraction techniques, we have discovered a quantum critical point in $Cd_2Os_2O_7$ as the all-in-all-out (AIAO) antiferromagnetic order is continuously suppressed to zero temperature and, concomitantly, the cubic lattice structure continuously changes from space group *Fd-3m* to *F-43m*. Surrounded by three phases of different time reversal and spatial inversion symmetries, the quantum critical region anchors two phase lines of opposite curvature, with striking departures from a mean-field form at high pressure. As spin fluctuations, lattice breathing modes, and quasiparticle excitations interact in the quantum critical region, we argue that they present the necessary components for strongly-coupled quantum criticality in this three-dimensional compound.

**TEXT:**

Fundamental symmetry constraints determine allowable collective states in solids [1]. Of particular recent interest is broken inversion symmetry in non-centrosymmetric materials, which can lead to hidden topological order and odd-parity superconductivity [1-3]. With the added consideration of broken time-reversal invariance, exotic magnetic states can emerge. A quantum critical point, where quantum fluctuations tie together competing ground states, is a fertile region to investigate and manipulate intertwined charge, spin, and structural degrees of freedom under different symmetry conditions. However, many experimental systems either manifest first-order quantum phase transitions without critical behavior, as exemplified by itinerant ferromagnets [4], or simply follow mean-field behavior [5,6]. Strongly-coupled systems with pronounced spin-orbit interactions provide a potential means to move to non-trivial criticality in three dimensions [4,5]. This scenario has been proposed for all-in-all-out (AIAO) antiferromagnetic order on a pyrochlore lattice [4], but remains to be identified experimentally.

The AIAO arrangement of spins on the pyrochlore lattice is an unusual form of magnetism that preserves the underlying cubic lattice symmetry. With strong spin-orbit coupling and low itinerant electron density, compounds with AIAO spin order are desirable candidates to explore non-trivial quantum critical behavior in three dimensions [5]. AIAO spin order has been observed in $FeF_3$, $Nd_2Zr_2O_7$, $A_2Ir_2O_7$ ($A$= Sm, Eu, and Nd), and $Cd_2Os_2O_7$ (Refs. 7-12) and suggested for additional $A_2Ir_2O_7$ systems with $A$=Y, Lu, Gd, Tb, Dy, Ho, and Yb [13]. For our purposes, $Cd_2Os_2O_7$ is most desirable. With the transition temperature ($T_N$=227K) roughly 60% higher than all other $A_2Ir_2O_7$ members, $Cd_2Os_2O_7$ should demonstrate the strongest correlation effects. Moreover, among spin-orbit coupled $5d$ compounds, only $Cd_2Os_2O_7$ and $Nd_2Ir_2O_7$ consistently manifest both an antiferromagnetic insulating phase and a metallic paramagnetic phase with $d\rho/dT$>0 (Refs. 13-15). It remains unclear whether iridates such as $Eu_2Ir_2O_7$ and $Sm_2Ir_2O_7$ are metallic or insulating in the paramagnetic phase, presumably due to vacancies and site disorder from the 3+/4+ valence condition [16]. By comparison, $Cd_2Os_2O_7$ exhibits less site disorder from its 2+/5+ valence condition, with no extraneous spin order arising from the $A$ site of the $A_2(Ir,Os)_2O_7$ structure [14].

Both pressure [14,17,18] and chemical tuning [15,18] of the $A$ site drive the insulating transition to lower temperature in the $A_2(Ir,Os)_2O_7$ compounds, but little is known about the behavior of the AIAO magnetic order. Here we directly explore the evolution of the spin, orbit, and lattice degrees of freedom, using polarization analysis to resonantly diffracted x-rays from $Cd_2Os_2O_7$ under diamond anvil cell pressures. This resonant diffraction technique is highly challenging with regard to both maintaining the single-crystal quality at cryogenic temperatures and tens of giga-Pascal pressures, and efficiently detecting weak magnetic diffraction signals from miniature samples of $5\times10^{-5}$ mm$^3$ size. Nevertheless, AIAO magnetic order and $5d$ orbital order can be resolved directly and tracked to the highest pressures under resonant conditions in the polarization switching π–σ channel of the (6, 0, 0) and (4, 2, 0) forbidden lattice orders, respectively [9]. The lattice symmetry and space group assignments were detected via the π–π' scattering channel.

**Results**
**The magnetic quantum phase transition.** We summarize in Fig. 1 the symmetry evolution under pressure. At ambient pressure, $Cd_2Os_2O_7$ has the pyrochlore structure of the *Fd-3m* space group (No. 227). Optical Raman scattering maps out a uniform *Fd-3m* phase up to $P$ = 29 GPa and between $T$ = 10 and 300 K. The phonon modes characteristic of that space group develop smoothly over the entire range (Fig. 2). The overall cubic symmetry is further verified by x-ray diffraction up to 41 GPa at 4 K (Fig. 3), as lineshapes of the (1, 1, 1), (2, 2, 2), (0, 2, 2), (0, 4, 4), and (4, 0, 0) diffraction orders remain single peak. The cubic lattice constant decreases smoothly under pressure, without discernable discontinuity at the quantum critical point $P_c$=35.8 GPa (Fig. 1, discussed below). While the *Fd-3m* space group is uniquely determined by the unit cell's lattice constant and one free coordinate $x$ for oxygen position on 48$f$ sites that characterizes the trigonal distortion of the $OsO_6$ cluster [14,19], diffraction intensities at (1, 1, 1) and (0, 2, 2) show a continuous evolution through the quantum phase transition, with $x$ increasing by a small amount from 0.319 at $P$=0 (Ref. 14) to ~0.325 at $P_c$ (Fig. 3b-c). Neither the AIAO antiferromagnetic order nor the continuous space group evolution within the cubic symmetry (Fig. 1) would be detectable by macroscopic approaches such as electrical transport. Instead, we address these issues using resonant single crystal diffraction with polarization analysis at the Os *L2* edge.

Resonant diffraction at both (6, 0, 0) and (4, 2, 0) in the π–σ and π–π' polarization channels are displayed in Fig. 4 for pressures that traverse the quantum phase boundary. The magnetic diffraction intensity $I_{(6, 0, 0)}$ in the π–σ channel at $E$=12.387 keV, which scales to the ordered staggered moment <$m$> as $I \sim $<$m$>$^2$, decreases continuously with increasing pressure (Fig. 5a). The scatter in the data makes it difficult to identify definitively the quantum critical region, but a phenomenological fit of the intensity data as $I\sim<m>^2\sim(P_c-P)^{2\beta}$ over the whole pressure range gives a critical pressure $P_c= 35.8\pm0.7$ GPa and an exponent $\beta=0.40\pm0.04$ for the order parameter $<m>$. From energetic considerations of localized 3$d$-5$d$ spins [20,21], our magnetic diffraction results at the low temperature limit of 4K also provide a means to estimate the AIAO magnetic phase boundary through $T_N\sim\Delta L\sim<m>^2$, where $\Delta L$ is the external magnetostriction that develops under $<m>$. In AIAO-ordered pyrochlores where the cubic lattice symmetry is preserved by the magnetic order and the phase transition is continuous, the external magnetostriction $\Delta L$ is difficult to observe over the lattice's thermal expansion. Nevertheless, a non-monotonic evolution of the lattice constant with temperature, $a(T)$, for AIAO order at very low $T$ has been demonstrated in $Nd_2Ir_2O_7$ ($T_N$=33.5K and $\Delta a/a\sim1$ $10^{-4}$) [22]. Following this logic, we identify the magnetic phase boundary $T_N(P)$ of $Cd_2Os_2O_7$ in Fig. 1 via the relationship $T_N\sim I\sim<m>^2$. At high pressure, for $P > P_c$, the presence of a charge resonance at (4, 2, 0) verifies that the unoccupied $t_{2g}$ orbitals remain the same in promoting the resonance behavior (Fig. 4e, 4f). At the same time, the absence of a resonance at (6, 0, 0) importantly marks the vanishing of the staggered moment $<m>$ and the long-range antiferromagnetic order (Fig. 4c).

**The lattice quantum phase transition.** With the disappearance of AIAO order at the magnetic critical point $P_c$, the charge diffraction starts to gain intensity in the $\pi-\pi'$ channel for both (6, 0, 0) and (4, 2, 0) (Fig. 4), and grows continuously for $P>P_c$ in an effectively exponential manner (Fig. 5b). We emphasize that these are forbidden diffraction orders in the *Fd-3m* space group. The energy dependence manifests no resonance, but is instead consistent with the inverted shape of the Os L2 absorption edge. At 41 GPa, 15% above $P_c$, there is no diffraction intensity at forbidden orders (3, 0, 0), (5, 0, 0), and (2, 1, 1). With preserved cubic symmetry, *F-43m* is the only maximal non-isomorphic subgroup of the original *Fd-3m* space group that is consistent with the measured diffraction results. We note that the pressure-induced AIAO quantum phase transition in $Cd_2Os_2O_7$ differs from the chemical tuning scenario of the $A_2Ir_2O_7$ series (*A* from Sm to Pr), where the lattice inversion symmetry is preserved throughout [13].

The space group *F-43m* breaks lattice inversion symmetry, opening possibilities for a number of exotic states. We did not detect further symmetry breaking such as the small tetragonal distortion in $Cd_2Re_2O_7$ that would remove the three-fold rotational symmetry [23]. The broken inversion symmetry derives from differently sized adjacent tetrahedra of the Os and Cd sublattices, which are no longer centrosymmetric and can be regarded as fully-softened breathing modes [3] with spontaneous symmetry breaking. These breathing modes disappear above $T_c \sim K|\Delta_{Os}|^2$, with the lower *F-43m* symmetry replaced by the higher *Fd-3m* symmetry with adjacent tetrahedra of equal size. Here $K$ is the vibrational elastic constant and $\Delta_{Os}$ is the amplitude of the Os tetrahedron breathing mode. We expect $T_c \sim I$, given that the measured x-ray diffraction intensities (Fig. 5b) depend on these lattice distortions as $I\sim|f_{Os}\Delta_{Os}+f_{Cd}\Delta_{Cd}|^2$, where $f_{Os/Cd}$ are x-ray atomic form factors. Hence $T_c$ increases two orders of magnitude in a short pressure interval: from approximately 5 K at 37 GPa to ~500 K at 41 GPa, leading to the striking concave high-pressure phase boundary in Fig. 1.

**Discussion**

The reduction of lattice symmetry from *Fd-3m* to *F-43m* in $Cd_2Os_2O_7$ has direct implications for phonon coupling [3] to spin fluctuations at the quantum critical point. Although spin-phonon coupling is known to generate unequal bond lengths in spin-Peierls dimers and antiferromagnetic superlattices [20], the breathing phonons are not fully softened in the AIAO phase and, at least in the static limit, AIAO order only induces an external magnetostriction with a homogenous expansion [22]. In the high-pressure phase, a breathing lattice could in principle still permit AIAO spin configurations on different sized tetrahedra, despite a loss of site inversion symmetry in the *F-43m* space group. Nevertheless, as no long-range commensurate antiferromagnetic order was observed experimentally (Fig. 4c), the magnetic ground state at high pressure is likely spin-disordered, as ferromagnetically interacting Ising moments along local <1,1,1> axes would generate a high level of frustration [7]. Tuning by either chemical doping or pressure drives the ratio of the magnetic interaction strength to the hopping integral smaller in $A_2$(Os,Ir)$_2$O$_7$. With an increasing electron density under 15% volume reduction by pressure, and moving away from the strong interaction strength limit [13], one would not naturally expect a ferromagnetic ground state. Furthermore, ferromagnetic quantum phase transitions, as well as commensurate-incommensurate antiferromagnetic transitions, are first order, which would contrast with the continuous AIAO quantum phase transition observed in our experiment.

The increased bandwidth under pressure suggests that the electronic properties of $Cd_2Os_2O_7$ in the high-pressure *F-43m* state become more metallic, potentially even superconducting in analogy to superconducting $Cd_2Re_2O_7$ with its broken inversion symmetry (Ref. 23). The relationship between the spin and charge transitions is also intriguing. AIAO magnetic order and the metal-insulator transition respond similarly to compression across the *P-T* phase diagram. Our projected magnetic phase boundary in Fig. 1 gives $dT_N/dP \sim -5.0$ K/GPa at $P=0$, which is consistent with $dT_{MIT}/dP = -4$ K/GPa in $Cd_2Os_2O_7$ measured over the first 2 GPa range [14]. Comparing the two 5*d* AIAO ordered compounds with clear high-temperature metallic states, we find an average $dT_N/dP \sim -6.5$ K/GPa over the whole AIAO phase in $Cd_2Os_2O_7$ and a $dT_{MIT}/dP \sim -5.8$ K/GPa in $Nd_2Ir_2O_7$ [17,18]. This comparison holds true despite large differences in $T_N$ of 227K and 33.5 K, respectively.

The experimentally observed coincidence of magnetic and structural phase transitions at $P_c = 35.8$ GPa, along with the similarities in the pressure evolution of the spin and charge degrees of freedom, point to one critical point within our experimental resolution (~1 GPa in the region around $P_c$) between the insulating AIAO order and the spin-disordered metallic phase. As Landau's phase transition theory would dictate two quantum phase transitions to be separated or to share a first-order phase line. If the transitions are separated, then our results point to a narrow intermediate region where the various spin, charge, and structural modes can couple across phases. If the phase boundaries actually meet at one critical point, then we may be in the realm of deconfined quantum criticality [24]. In any case, the concurrence of magnetic and structural phase transitions within our experimental resolution, and the clear deviation from mean-field behavior of the high-pressure phase, stimulate a discussion of strong coupling in the quantum critical region.

From the band structure perspective, the Os 5*d* $t_{2g}$ band in $Cd_2Os_2O_7$ is neither degenerate and forming a $S=3/2$ state under Hund's rule, as indicated by the reduced staggered moment $<m>=0.59$ $\mu_B$/Os, nor cleanly separated into several narrow bands as demonstrated for a perfect $OsO_6$ octahedron [14,25,26]. Instead, the $t_{2g}$ orbitals in $Cd_2Os_2O_7$ extend continuously over a spectral width of order 2 eV from the combined effect of $U$ (~1eV) [27], spin-orbit coupling (~0.35eV) [26], and trigonal distortion (~0.3eV) on the $OsO_6$ octahedron [19]. Through the continuous quantum phase transition, the overall stability of the empty $t_{2g}$ band is verified by the constant charge resonance profile at (4, 2, 0), with a coarse energy resolution slightly above 1 eV. From a metallic paramagnetic state, the formation of antiferromagnetic order would influence the oscillating dynamic component of the quasiparticle self-energy [28], and in turn introduce a spin-order dependent repulsion between empty and filled states that could account for the insulating state below $P_c$. This effective exchange field due to neighboring opposite spins is a generalized Slater mechanism, even without introducing Brillouin zone folding [28]. If the gap opens through a critical state of singular points at the Fermi surface, instead of a removal of states altogether, its thermodynamics could fit the Lifshitz description [27,29].

The essentially concurrent, continuous quantum phase transitions of antiferromagnetism, structure, and (apparently) charge in $Cd_2Os_2O_7$ provides the necessary ingredients for a generic approach to strongly-coupled, non-mean-field quantum criticality in three-dimensions [5]. With the Fermi surface fully gapped, quasiparticle fluctuations would involve all itinerant states in reducing the screening on Coulomb $U$, and the increased interaction range would then help stabilize a continuous quantum phase transition [5]. Indeed, at the ambient-pressure metal-insulator transition in $Cd_2Os_2O_7$, an increase in $U$ from 0.8 to 1.5 eV in the theoretical modeling is consistent with the observed spectral weight shift in infrared conductivity over the broad range of 0-4 eV (Ref. 30). Above the quantum critical point and in the spin-disordered phase space of the *Fd-3m* space group, soft AIAO spin fluctuations and lattice breathing modes could exist and compete, and further couple to quasiparticle fluctuations. The competition between spin and lattice degrees of freedom might explain the remarkable concave-shaped phase line at high pressure, as $T_c$ scales to pressure with a non-trivial exponent much larger than one, a characteristic of strongly-coupled quantum criticality [5]. We note as well that the quantum critical region is asymmetric in *P-T* phase space, as the magnetic and structural phase lines approach $P_c$ with different asymptotic behavior.

The quantum phase transition in $Cd_2Os_2O_7$, with its interwoven spin, orbit, lattice, and charge degrees of freedom, contrasts sharply with systems that have a partially gapped Fermi surface, exemplified by itinerant spin density waves where persistent carriers screen spin fluctuations and lead to mean-field behavior [4-6,31]. The cubic AIAO antiferromagnet also differs from itinerant ferromagnets, where strong spin and charge mode coupling at wave vector ***q***=0 categorically induces first-order quantum phase transitions [4]. Spin-orbit coupling in *5d* systems is regarded as intermediately strong [13], and pressure drives $U/t$ smaller with increasing kinetic energy $t$, away from the strong-correlation limit. Pressure tuning thus likely induces a continuous quantum phase transition while still preserving non-trivial quantum criticality in this *5d* antiferromagnet. By Luttinger's theorem, a continuous insulator-metal transition would result in either a carrier-mass enhancement or non-Fermi liquid behavior [32] in $Cd_2Os_2O_7$'s high-pressure phase, and the broken inversion symmetry sets the conditions for odd-parity superconductivity. A microscopic theory remains to be developed to describe the interaction between the AIAO spin fluctuations, breathing phonon modes, and quasiparticle excitations, especially taking into consideration the symmetry, chirality, and wave vector characteristics of each.

**Methods:**
**Resonant X-ray diffraction under high-pressure.** Both charge orbital order and magnetic order can induce resonant behavior at an absorption edge in x-ray diffraction [9,12,31]. At the Os $L_2$ edge, these two types of resonances share the empty part of the $t_{2g}$ band as the intermediate state, and exhibit the same resonance profiles. The charge resonance originates from the anisotropic tensor susceptibility. In general, it can be observed at many forbidden lattice orders and, specifically, at both (4, 2, 0) and (6, 0, 0) in $Cd_2Os_2O_7$. To avoid the charge anisotropic tensor susceptibility resonance and to isolate the magnetic resonance of the AIAO order, x-ray diffraction of the (6, 0, 0) order in $Cd_2Os_2O_7$ was performed with a limited azimuthal angle range around 45° relative to the (0, 0, 1) vector, using a horizontal diffraction geometry for both π-σ and π-π' polarization analyses [9].

For the diamond anvil cell high-pressure environment [33], we used a transmission (Laue) diffraction geometry in contrast to the typical reflection (Bragg) geometry at ambient pressure [9]. To achieve the diffraction geometry with the specified azimuthal condition at (6, 0, 0), single crystal samples were prepared in thin plate form with a surface normal along (0, -1, 1). This allows access to other diffraction orders such as (1, 1, 1), (4, 0, 0), (0, 2, 2) and (4, 2, 0) within the confined diamond anvil cell geometry, with the forbidden order (4, 2, 0) providing access to the charge-based anisotropic tensor susceptibility resonance. Several plates were polished down to 13 μm thickness, equivalent to one absorption length in $Cd_2Os_2O_7$ for x-rays at the Os $L2$ edge ($E$=12.387keV). Unlike the iridates, resonance behavior at both the $L2$ and $L3$ edges are present in osmates. We chose the resonance at the $L2$ edge at high $P$ because the higher energy x-rays had a longer x-ray penetration length through both the pressure environment and the sample.

The low-temperature, high-pressure diffraction setup at beamline 4-ID-D of the Advanced Photon Source has been described elsewhere [33]. To reduce absorption and enhance the signal-to-background ratio, a pair of wide-angle perforated Boehler diamond anvils [33] (SYNTEK Co. LTD., Japan) were used, with culet size varying from 800 to 550 μm. A methanol/ethanol 4:1 mixture was used as the pressure medium inside rhenium gaskets. Pressure was calibrated by a Ag manometer *in situ* at 4.0±0.5K using a two-parameter Birch equation of state, with $B_0$=108.85 GPa and $B$'=d$B$/d$P$=5.7 over the large pressure range. For x-ray polarization analysis, a highly oriented pyrolytic graphite (HOPG) plate of 5 mm thickness and 0.35° FWHM mosaic was used as the polarization analyzer. The (0, 0, 10) diffraction order of graphite at the Os $L2$ edge of 12.387 keV introduces a leakage of approximately 1.3% of the intensity from the π-π' channel to the π-σ channel, and vice versa. Data presented here were collected from a total of 8 samples under pressure for spin (7) and charge (4) resonances. The absence of high-pressure commensurate antiferromagnetic order was verified on two crystals.

**Optical Raman scattering.** Shards of single crystal $Cd_2Os_2O_7$ with original growth surfaces were individually loaded with a Neon pressure medium in a diamond anvil cell, and subsequently thermally cycled in a liquid helium flow cryostat. Optical Raman scattering was performed using a Horiba LabRam HR Evolution system in the MRSEC facilities at the University of Chicago, equipped with a 633nm wavelength laser for excitation.

**Data availability.** The data that support the findings of this study are available from the corresponding authors upon request.

**Acknowledgments**
We are grateful for stimulating discussions with J. Alicea, D. Belitz, L. Hozoi, J.C. Lang, P.A. Lee, G. Rafael, N. Shannon, O. Tchernyshyov, and M. Van Veenendaal. We thank J. Jureller for providing access to the optical Raman system and S. Tkachev at GSECARS for help on high-pressure Neon loading. The use of shared facilities of the University of Chicago Materials Research Science and Engineering Center (MRSEC) was supported by National Science Foundation Grant No. DMR-1420709. Use of the COMPRES-GSECARS gas loading system was supported by COMPRES under NSF Cooperative Agreement EAR -1606856 and by GSECARS through NSF grant EAR-1634415 and DOE grant DE-FG02-94ER14466. The work at Caltech was supported by National Science Foundation Grant No. DMR-1606858. The work at the Advanced Photon Source of Argonne National Laboratory was supported by the US Department of Energy Basic Energy Sciences under Contract No. NEAC02-06CH11357. D.M. acknowledges support from the U.S. Department of Energy, Office of Science, Basic Energy Sciences, Materials Sciences and Engineering Division.

**Author contributions**
Y.F. and T.F.R. conceived of the research. D.M. provided single crystal samples. Y.W., Y.F., A.P., Y.R., and J.-W. K. performed x-ray measurements. Y.W., Y.F., and T.F.R. analyzed the data and prepared the manuscript. All authors commented on the manuscript.

**Additional information**
Correspondence and requests for materials should be addressed to T.F.R. < tfr@caltech.edu > or Y.F. < yejun@oist.jp >.

**Competing financial interests**
The authors declare no competing financial interests.

**Figure captions:**
**Fig. 1 | Symmetries across the *P-T* phase diagram.** The $Cd_2Os_2O_7$ lattice retains its cubic symmetry throughout the probed *P-T* phase space, but continuously transitions between *Fd-3m* and *F-43m* space groups. The *Fd-3m* lattice symmetry was verified by optical Raman scattering from 0-29 GPa and 10-300 K (grey crosses), while both phases of magnetism and structure (pink and blue shading) were inferred from x-ray diffraction measurements at $T = 4$ K. The metallic paramagnetic phase in the low-pressure *Fd-3m* space group has both spatial inversion ($\mathcal{I}$) and time reversal ($\mathcal{T}$) symmetries. For the spin degrees of freedom, time reversal symmetry is broken in the low-pressure AIAO phase. On the high-pressure side, the *F-43m* space group breaks the spatial inversion symmetry, introduces a tetrahedral breathing distortion (inset), and restores the time reversal symmetry with disordered spins. The coincidence of continuous magnetic and structural phase transitions at a single quantum critical point suggests strong coupling between spin, orbit, lattice, and (potentially) charge degrees of freedom in this *5d* pyrochlore compound.

**Fig. 2 | Optical Raman scattering.** Raw Raman spectra of $Cd_2Os_2O_7$ at various temperatures from 10 to 300 K and our highest measured pressure, $P \sim 29$ GPa. We only observe the six modes consistent with the *Fd-3m* space group. The $T_{2g}(3)$ and $A_{1g}$ modes merge under their broadened peaks.

**Fig. 3 | Lattice evolution under pressure.** **(a)** Pressure evolution of the lattice constant was fit to a two-parameter Birch equation of state with $B$=190.4±3.6 GPa, and $B'$ = 4.2±0.2. (insets) Longitudinal ($\theta/2\theta$) scans of (1, 1, 1) and (0, 2, 2) orders measured at various pressures using 12.387 keV x-rays verify the cubic symmetry. **(b-c)** Pressure evolution of integrated diffraction intensities of (0, 2, 2) and (1, 1, 1) orders, normalized by (0, 4, 4) and (2, 2, 2) orders, respectively. The measurement was performed under either resonant ($E$=12.387 keV) or off-resonant ($E$=12.355 keV) conditions. These two orders are sensitive to O 48*f* sites in the unit cell, and develop in opposite fashion up to 40 GPa. **(d)** Simulated (0, 2, 2) and (1, 1, 1) intensities as a function of *x*. The overall percentage changes of (0, 2, 2) and (1, 1, 1) give an *x* increasing from 0.319 at *P*=0 (Ref. 14) to approximately 0.325 at $P_c$.

**Fig. 4 | Polarization-sensitive resonant diffraction data under pressure.** Raw energy scan profiles at both **(a-c)** (6, 0, 0) and **(d-f)** (4, 2, 0) orders from two separate polarization channels (π-σ in red/pink and π-π' in navy/aqua), expressed in counts/second for 100 mA synchrotron

storage ring current. The azimuthal angel ψ relative to the (0, 0, 1) vector is specified for each order. At (6, 0, 0), the π-σ channel manifests magnetic resonant spectra whose intensity decreases continuously with pressure. Beyond 36 GPa, the sharp resonance disappears, and a background similar to the Os *L2* absorption edge shape remains (red, Fig. 3c), which comes from x-ray fluorescence that passed through the polarization analyzer. Its origin is demonstrated by a comparison with an energy scan at (5.98, 0,0) (gray, Fig. 3c), where no magnetic diffraction is expected, and also with an Os *L2* absorption spectrum at 39.3 GPa (black, Fig. 3c). The fluorescence part of the background intensity is dependent on the detector slit size. At (4, 2, 0), the π-σ channel reveals the orbital ordering via the anisotropic tensor susceptibility resonance. For both (6, 0, 0) and (4, 2, 0) orders, the π-π' charge diffraction intensities rise dramatically at pressures above the magnetic phase boundary, with a small leakage into the π-σ channel becoming apparent in Fig. 3e through the polarization analyzer (Methods).

**Fig. 5 | Continuous magnetic and structural quantum phase transitions.** (a) Magnetic diffraction intensity was measured at (6, 0, 0) and in the π−σ channel, with a power law fit (solid line) to model the evolution over the whole pressure range. (b) Lattice diffraction intensities, measured at both the (6, 0, 0) and (4, 2, 0) orders in the π-π' channel, indicate a continuous switching between the *Fd-3m* and *F-43m* space groups with a phase boundary that rises effectively exponentially. All intensities were measured at 4.0±0.5 K with integration of sample mosaic profiles using 12.387 keV x-rays.

**References:**

[1] Soluyanov, A.A. *et al*., Type-II Weyl semimetals. *Nature* **527**, 495-498 (2015).

[2] Harter, J.W., Zhao, Z.Y., Yan, J.-Q., Mandrus, D.G., & Hsieh, D. A parity-breaking electronic nematic phase transition in the spin-orbit coupled metal $Cd_2Re_2O_7$. *Science* **356**, 295-299 (2017).

[3] Bzdusek, T., Rüegg, A. & Sigrist, M. Weyl semimetal from spontaneous inversion symmetry breaking in pyrochlore oxides. *Phys. Rev. B* **91**, 165105 (2015).

[4] Belitz, D., Kirkpatrick, T. R. & Vojta, T. How generic scale invariance influences quantum and classical transitions, *Rev. Mod. Phys.* **77**, 579-632 (2005).

[5] Savary, L., Moon, E.-G. & Balents, L. New type of quantum criticality in pyrochlore iridates. *Phys. Rev. X* **4**, 041027 (2014).

[6] Feng, Y. *et al*. Itinerant density wave instabilities at classical and quantum critical points. *Nat. Phys.* **11**, 865-871 (2015).

[7] Bramwell, S. T. & Harris, M.J. Frustration in Ising-type spin models on the pyrochlore lattice. *J. Phys.: Condens. Matter* **10**, L215-L220 (1998).

[8] Tomiyasu, K. *et al*., Emergence of magnetic long-range order in frustrated pyrochlore $Nd_2Ir_2O_7$ with metal–insulator transition. *J. Phys. Soc. Jpn*. **81**, 034709 (2012).

[9] Yamaura, J. *et al*. Tetrahedral magnetic order and the metal-insulator transition in the pyrochlore lattice of $Cd_2Os_2O_7$. *Phys. Rev. Lett*. **108**, 247205 (2012).

[10] Sagayama, H. *et al*., Determination of long-range all-in-all-out ordering of $Ir^{4+}$ moments in a pyrochlore iridate $Eu_2Ir_2O_7$ by resonant x-ray diffraction. *Phys. Rev. B* **87**, 100403 (2013).

[11] Lhotel, E. *et al*., Fluctuations and all-in-all-out ordering in dipole-octupole $Nd_2Zr_2O_7$. *Phys. Rev. Lett.* **115**, 197202 (2015).

[12] Donnerer, C. *et al*. All-in–all-out magnetic order and propagating spin waves in $Sm_2Ir_2O_7$. *Phys. Rev. Lett*. **117**, 037201 (2016).

[13] Witczak-Krempa, W., G. Chen, G., Kim, Y. B. & Balents, L. Correlated quantum phenomena in the strong spin-orbit regime, *Annu. Rev. Condens. Matter Phys*. **5**, 57-82 (2014).

[14] Mandrus, D. *et al*. Continuous metal-insulator transition in the pyrochlore $Cd_2Os_2O_7$. *Phys. Rev. B* **63**, 195104 (2001).

[15] Matsuhira, K., Wakeshima, M., Hinatsu, Y. & Takagi, S. Metal-insulator transitions in pyrochlore oxides $Ln_2Ir_2O_7$, *J. Phys. Soc. Jpn*. **80**, 094701 (2011).

[16] Ishikawa, J. J., O'Farrell, E. C. T. & Nakatsuji, S. Continuous transition between antiferromagnetic insulator and paramagnetic metal in the pyrochlore iridate $Eu_2Ir_2O_7$. *Phys. Rev. B* **85**, 245109 (2012).

[17] Sakata, M. *et al*. Suppression of metal-insulator transition at high pressure and pressure-induced magnetic ordering in pyrochlore oxide $Nd_2Ir_2O_7$. *Phys. Rev. B* **83**, 041102 (2011).

[18] Ueda, K., Fujioka, J., Terakura, C. & Tokura, Y. Pressure and magnetic field effects on metal-insulator transitions of bulk and domain wall states in pyrochlore iridates, *Phys. Rev. B* **92**, 121110 (2015).

[19] N. A. Bogdanov, R. Maurice, I. Rousochatzakis, J. van den Brink, L. Hozoi, Magnetic state of pyrochlore $Cd_2Os_2O_7$ emerging from strong competition of ligand distortions and longer-range crystalline anisotropy. *Phys. Rev. Lett*. **110**, 127206 (2013).

[20] Feng, Y., Jaramillo, R., Banerjee, A., Honig, J. M. & Rosenbaum, T.F. Magnetism, structure, and charge correlation at a pressure-induced Mott-Hubbard insulator-metal transition. *Phys Rev. B* **83**, 035106 (2011).

[21] Wang, Y. *et al*. Spiral magnetic order and pressure-induced superconductivity in transition metal compounds, *Nat. Commun*. **7**, 13037 (2016).

[22] Takatsu, H., Watanabe, K., Goto, K. & Kadowaki, H. Comparative study of low-temperature x-ray diffraction experiments on $R_2Ir_2O_7$ (R=Nd, Eu, and Pr). *Phys. Rev. B* **90**, 235110 (2014).

[23] Yamaura, J.I. & Hiroi, Z. Low temperature symmetry of pyrochlore oxide $Cd_2Re_2O_7$. *J. Phys. Soc. Jpn.* **71**, 2598-2600 (2002).

[24] Senthil, T., Vishwanath, A., Balents, L., Sachdev, S. & Fisher, M.P.A. Deconfined quantum critical points. *Science* **303**, 1490-1494 (2004).

[25] Calder, S. *et al*. Spin-orbit-driven magnetic structure and excitation in the 5*d* pyrochlore $Cd_2Os_2O_7$. *Nat. Commun.* **7**, 11651 (2016).

[26] Taylor, A. E. *et al*. Spin-orbit coupling controlled *J* =3/2 electronic ground state in 5$d^3$ oxides. *Phys. Rev. Lett*. **118**, 207202 (2017).

[27] Shinaoka, H., Miyake T. & Ishibashi, S. Noncollinear magnetism and spin-orbit coupling in 5*d* pyrochlore oxide $Cd_2Os_2O_7$. *Phys. Rev. Lett*. **108**, 247204 (2012).

[28] Arita, R., Kunes, J., Kozhevnikov, A.V., Eguiluz, A. G. & Imada, M. *Ab initio* studies on the interplay between spin-orbit interaction and Coulomb correlation in $Sr_2IrO_4$ and $Ba_2IrO_4$. *Phys. Rev. Lett*. **108**, 086403 (2012).

[29] Lifshitz, I. M. Anomalies of electron characteristics of a metal in the high pressure region. *Sov. Phys. JETP* **11**, 1130-1135 (1960).

[30] Sohn, C.H. *et al*. Optical spectroscopic studies of the metal-insulator transition driven by all-in–all-out magnetic ordering in 5*d* pyrochlore $Cd_2Os_2O_7$. *Phys. Rev. Lett*. **115**, 266402 (2015).

[31] Feng, Y. *et al*. Incommensurate antiferromagnetism in a pure spin system via cooperative organization of local and itinerant moments. *Proc. Nat. Acad. Sci. USA* **110**, 3287-3292 (2013).

[32] Imada, M., Fujimori, A. & Tokura, Y. Metal-insulator transitions. *Rev. Mod. Phys.* **70**, 1039-1263 (1998).

[33] Feng, Y. *et al*. Hidden one-dimensional spin modulation in a three-dimensional metal, *Nat. Commun*. **5,** 4218 (2014).

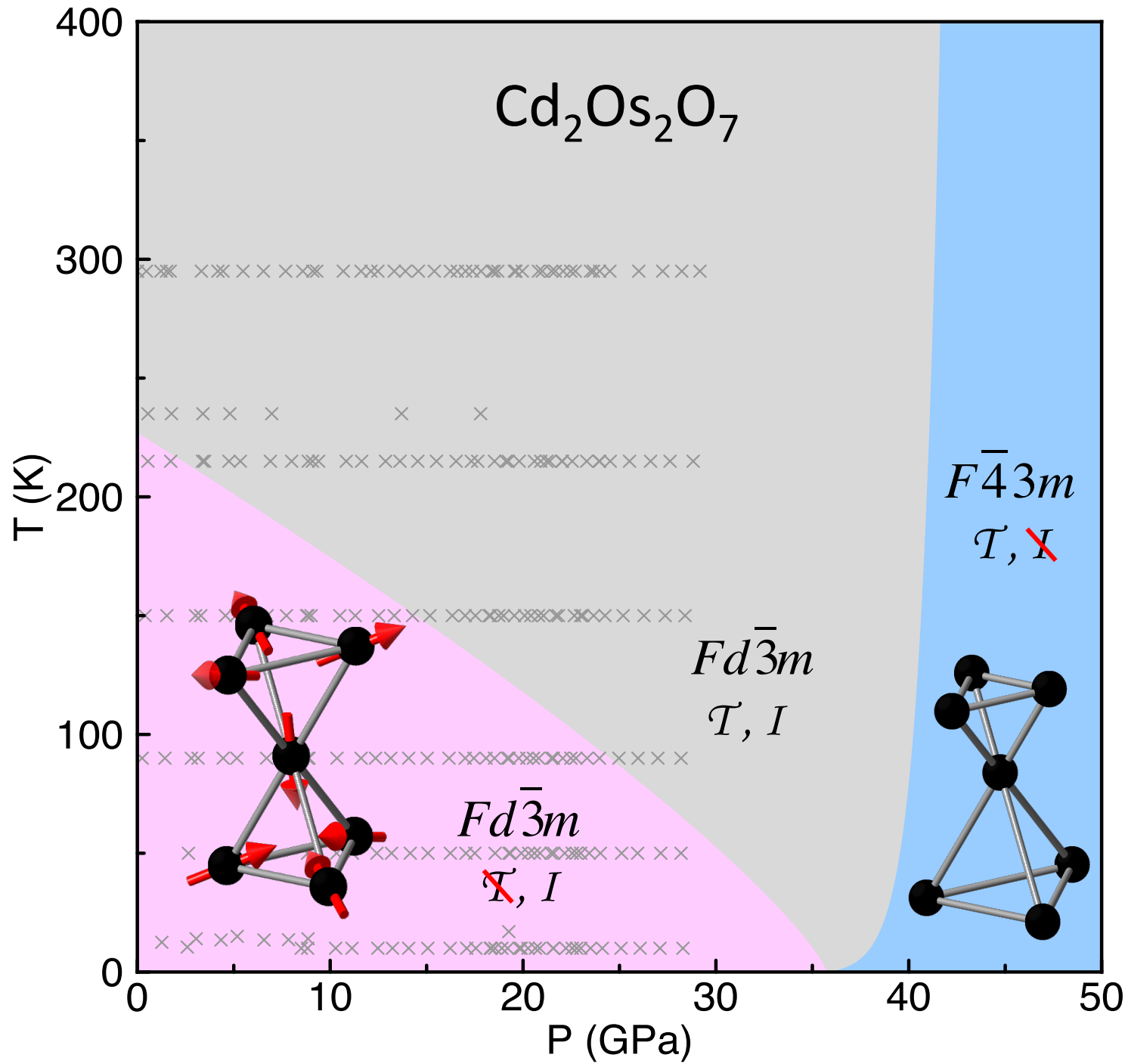

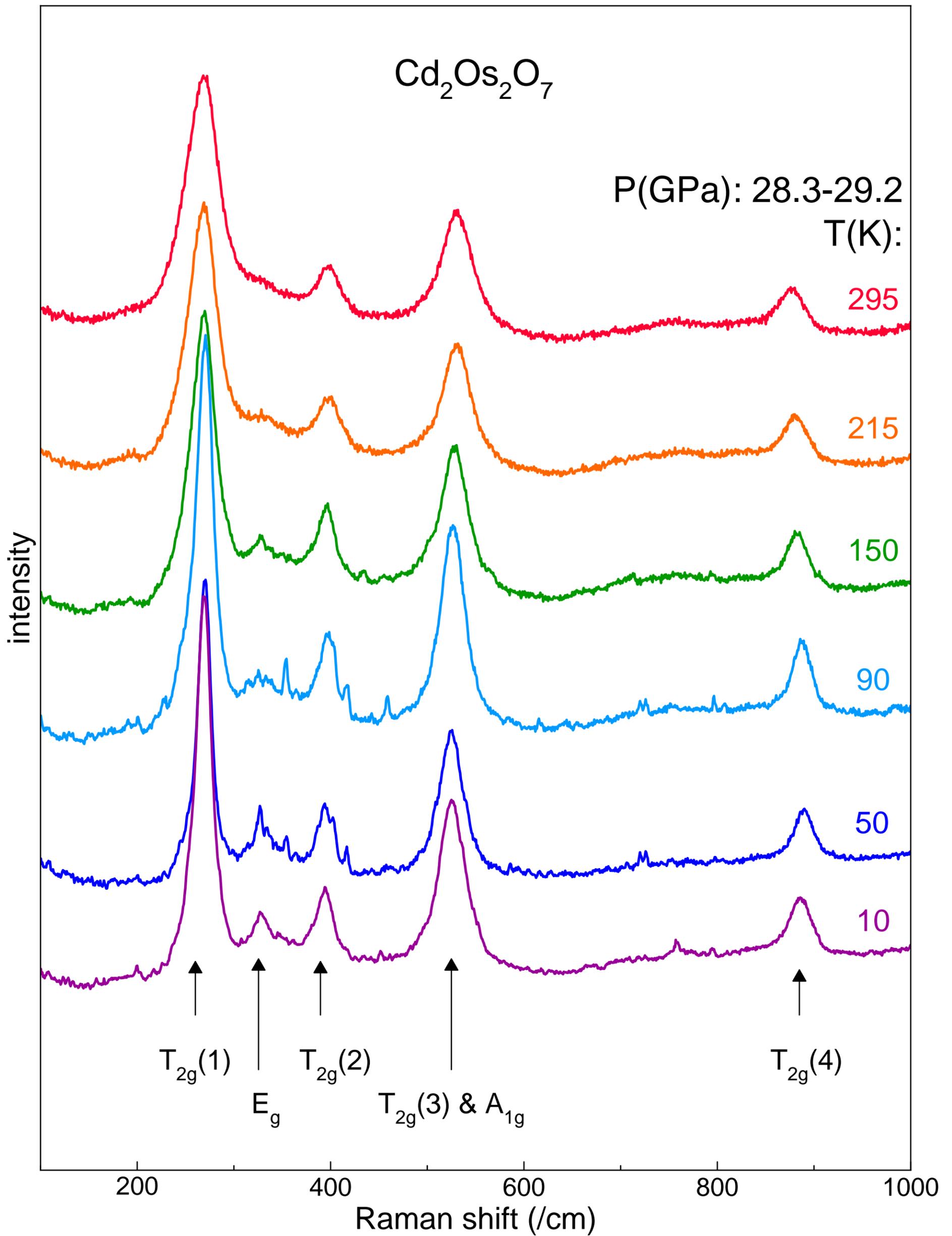

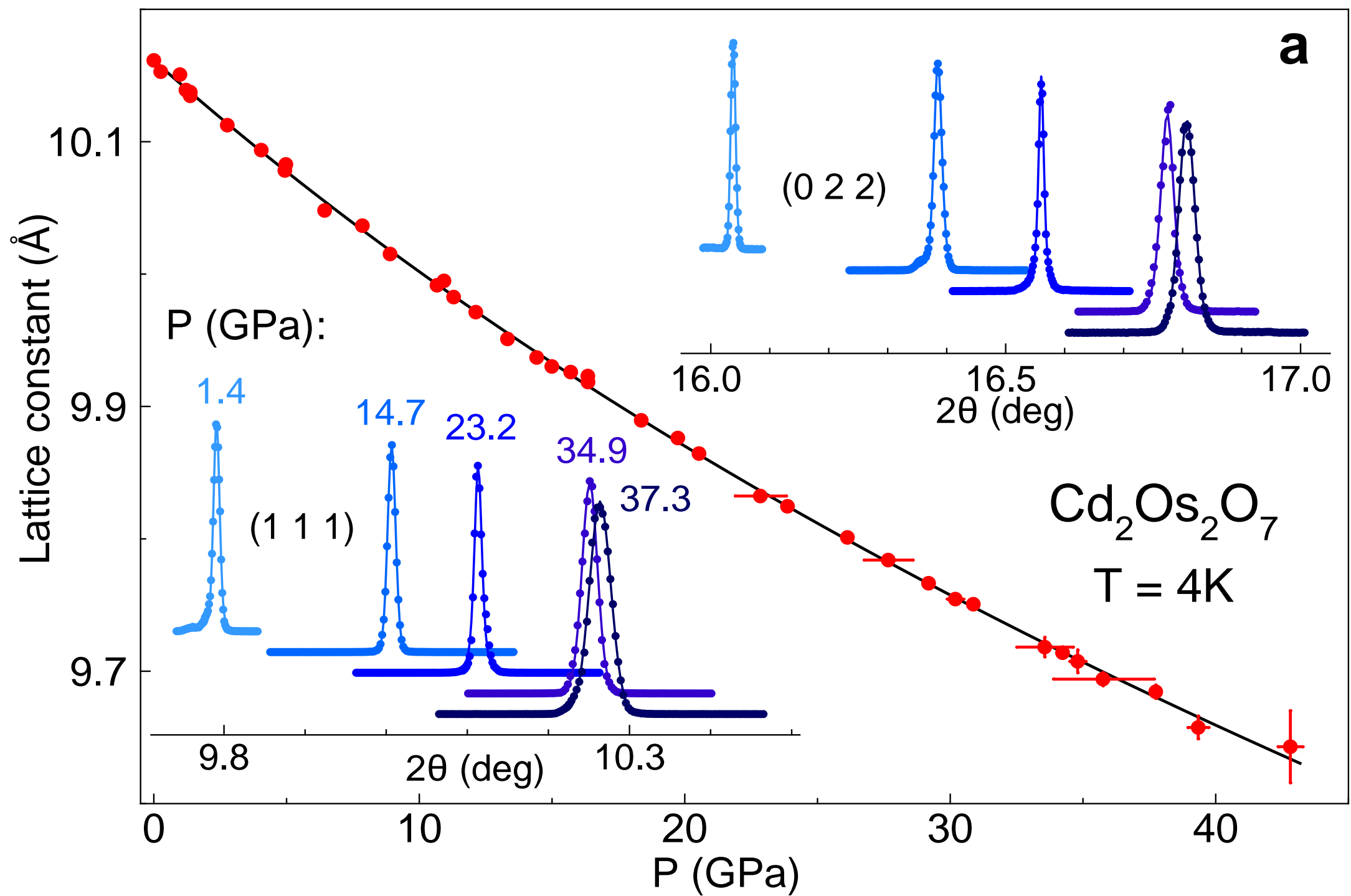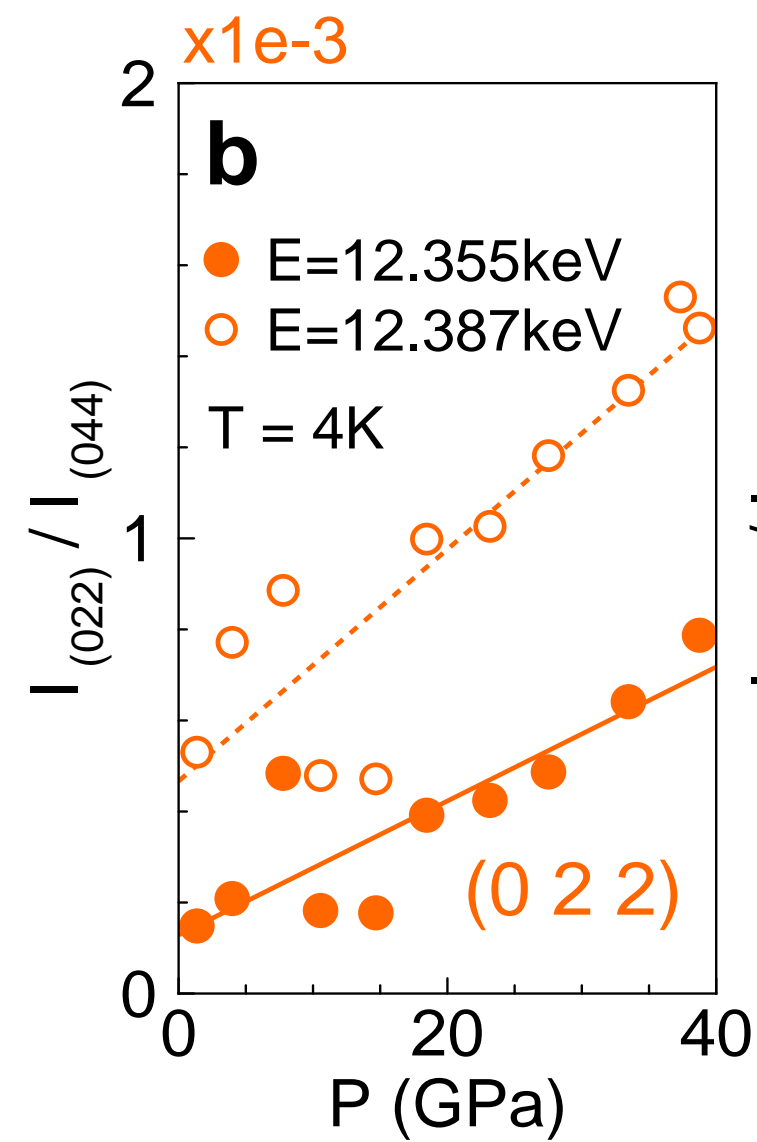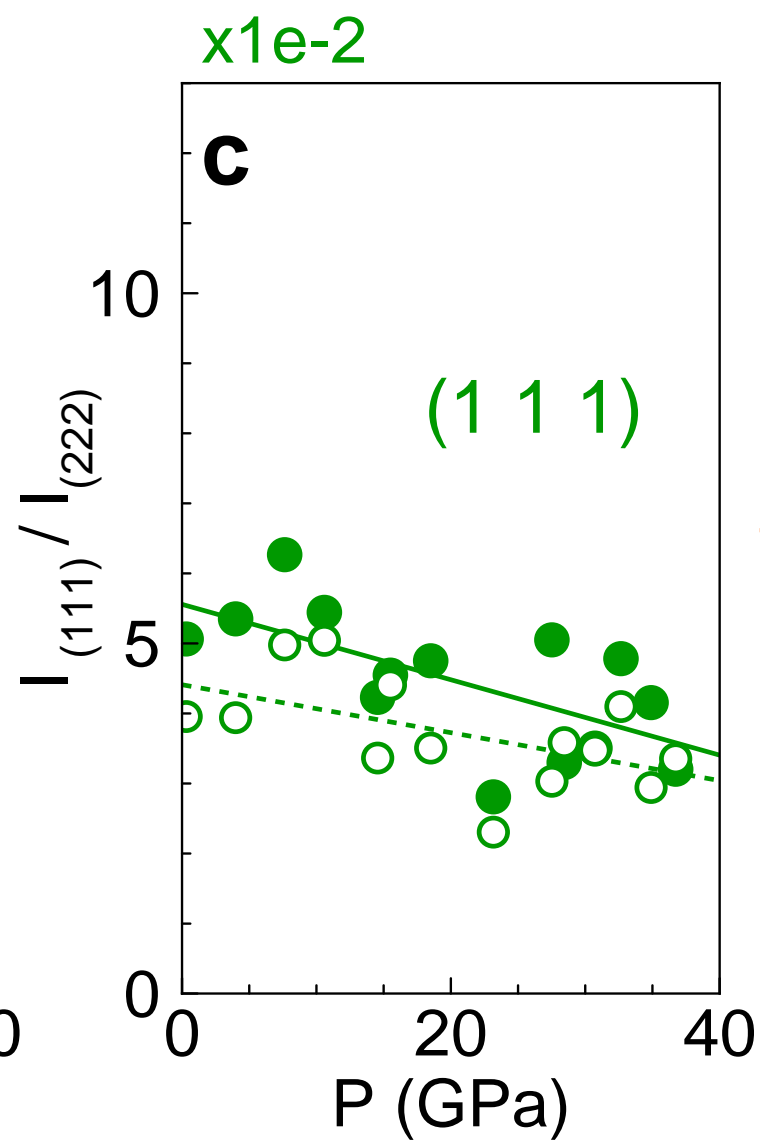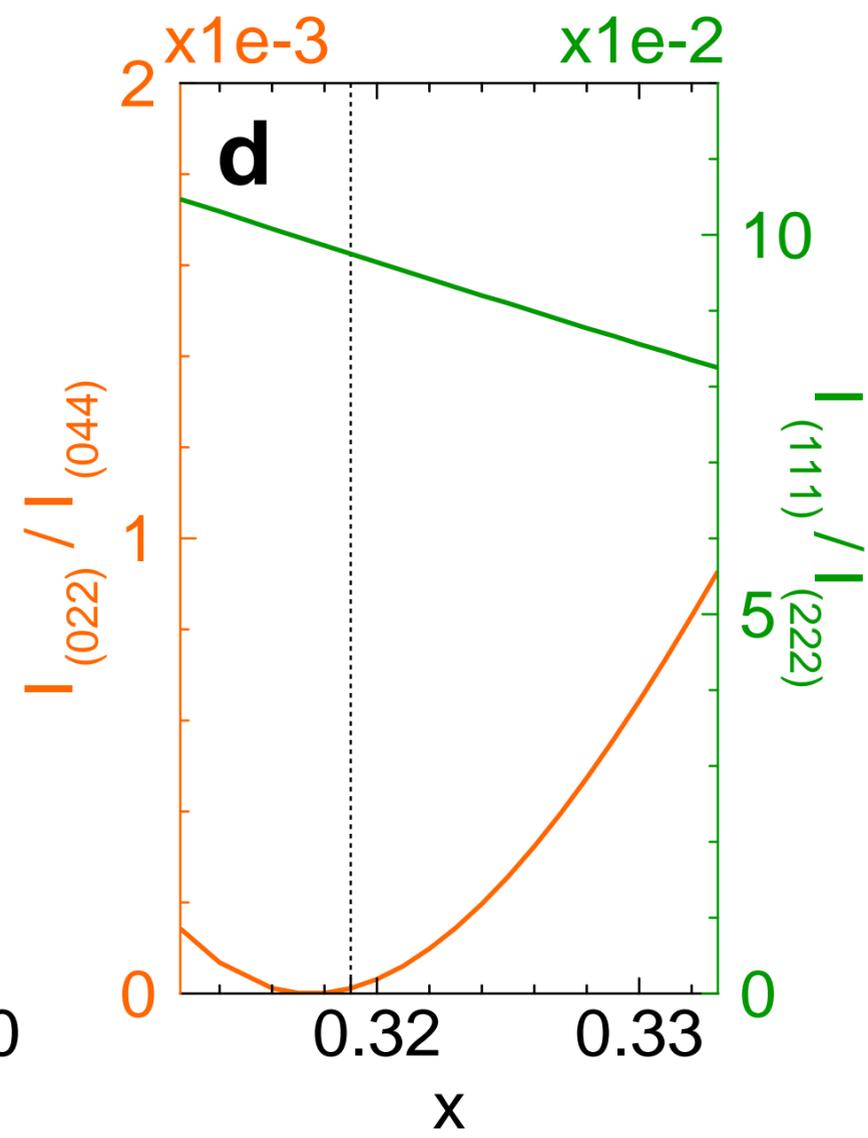

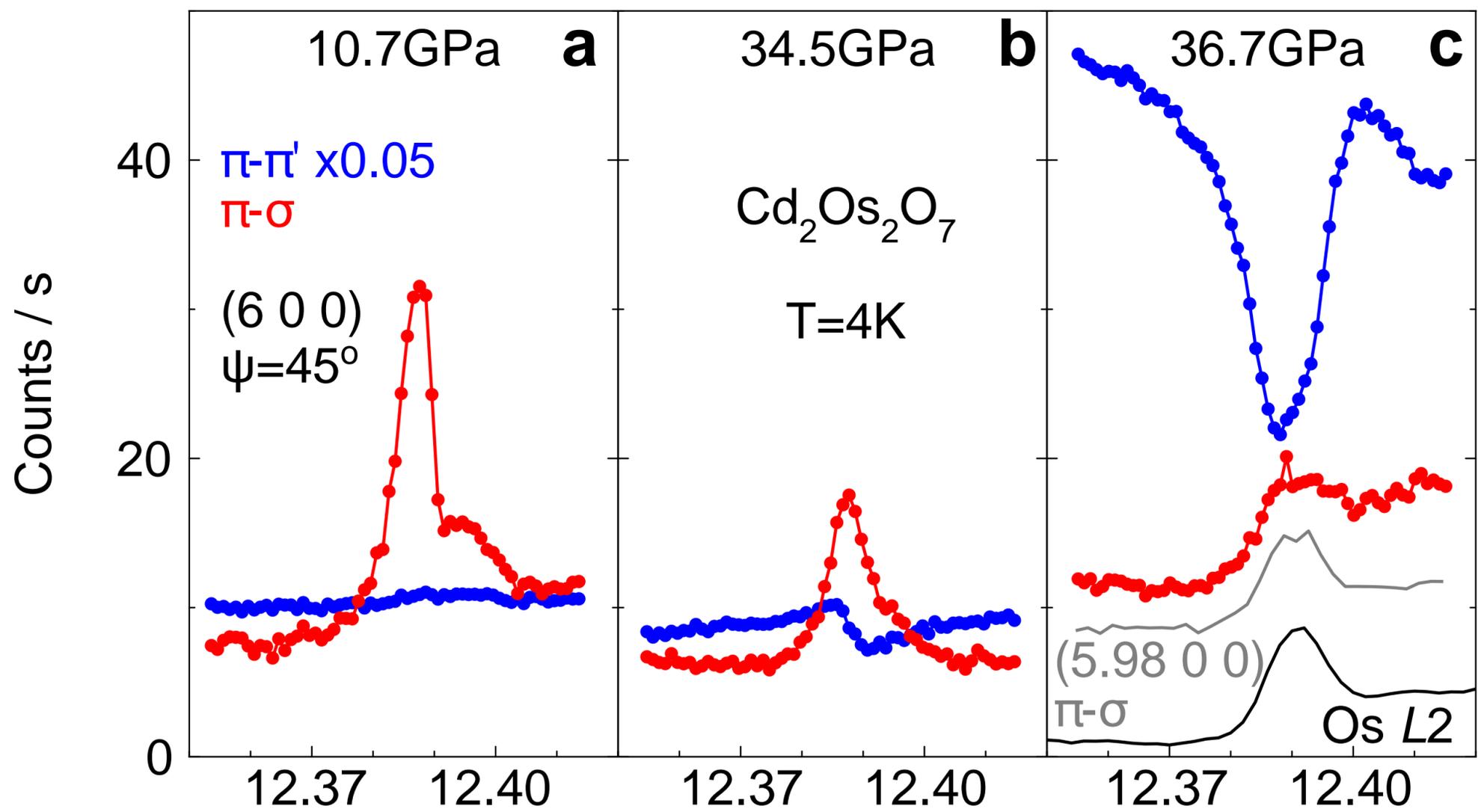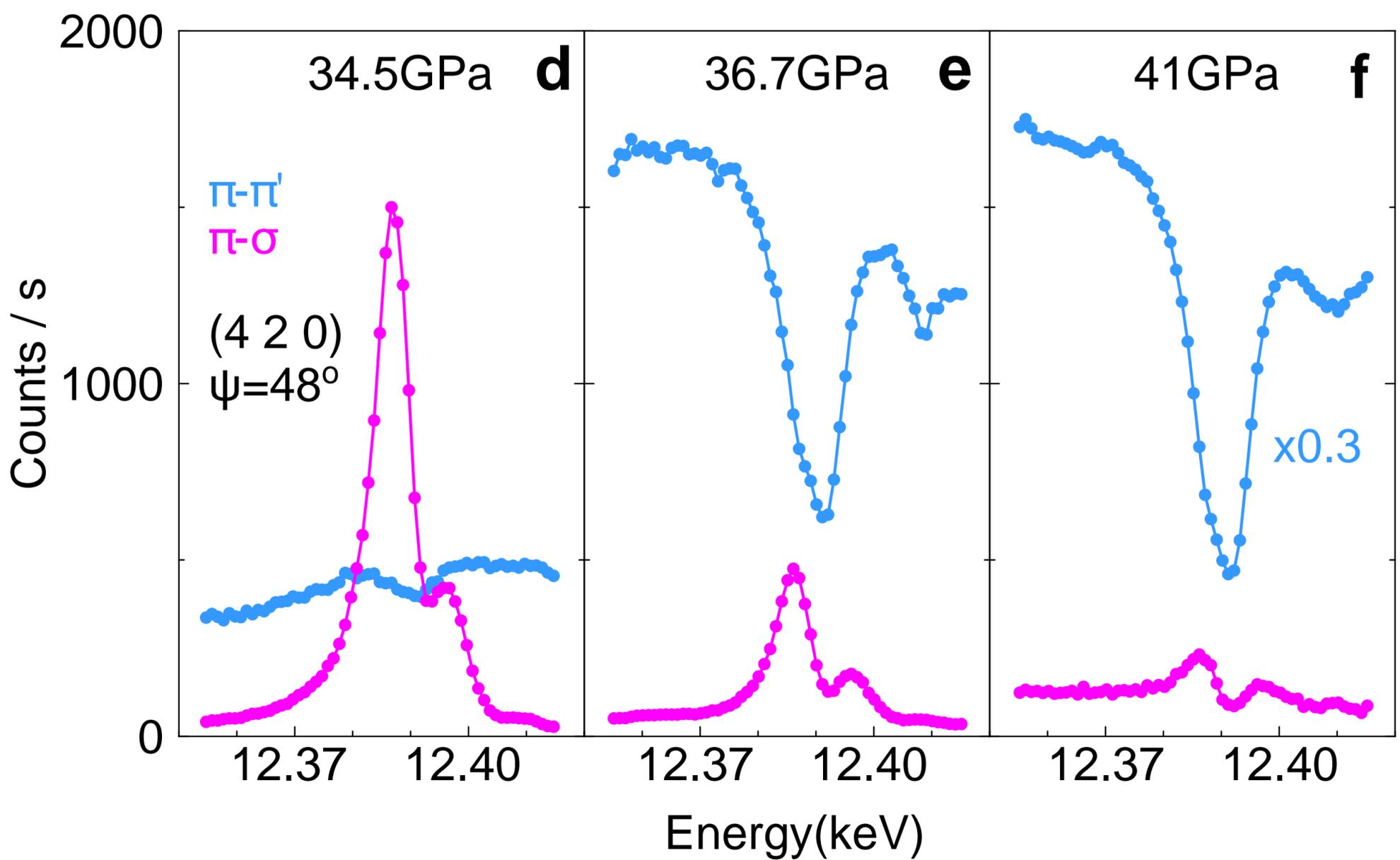

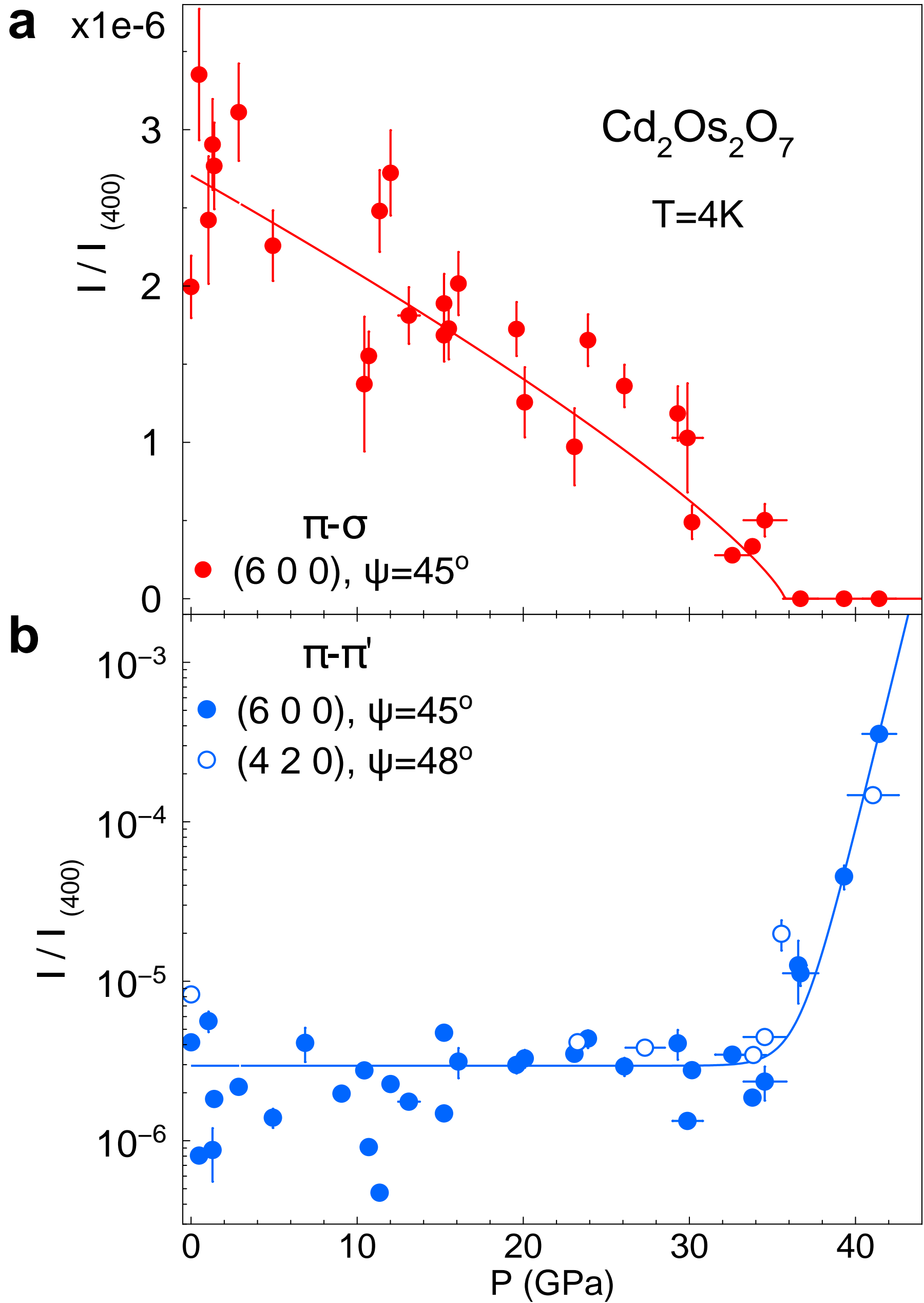